\documentclass[aps,pre,twocolumn,10pt,
               superscriptaddress,longbibliography,
               amsmath,amsfonts]{revtex4-2}

\usepackage{mathrsfs,graphicx,comment,xcolor,physics}

\usepackage{hyperref}  
\hypersetup{colorlinks,citecolor=red,linkcolor=blue,
            filecolor=black,urlcolor=black}

\begin{document}
\title{Retained hidden excess generates memory in price-limited markets}
\author{Debraj Das}
\email{ddas@sissa.it}
\affiliation{SISSA -- International School for Advanced Studies, Via Bonomea 265, 34136 Trieste, Italy}
\affiliation{Istituto dei Sistemi Complessi, Consiglio Nazionale delle Ricerche, via Madonna del Piano 10, 50019 Sesto Fiorentino, Italy}

\begin{abstract}
The daily return of a stock is often restricted to an exchange-imposed band to curb extreme fluctuations. Any attempted price movement beyond this band is clipped, leaving an unobserved excess. We introduce a minimal stochastic latent-state model in which a fraction of this hidden excess is retained for the next day. This retention generates memory, even though the daily stochastic driving shocks are independent. For symmetric driving shocks with regularly varying tails, the stationary latent return preserves the tail index of the noise, but has an enhanced tail amplitude. In the wide-band limit,  a close of the daily return at either limit of the band admits a single-dominant-shock description. We show that after such an event, the mean return on the following day has the same sign and grows proportionally to the band width, while the probability of reaching the same limit again approaches a finite value. Reaching the opposite band limit on the following day requires a second extreme shock of opposite sign and is power-law suppressed. Simulations support these analytical predictions. Empirical data from stocks subject to daily price limits are qualitatively consistent with the predicted same-sign response and its increase across wider price bands.
\end{abstract}
\maketitle

\section{Introduction}

Rare and extreme fluctuations in correlated stochastic
systems are a central topic in statistical physics~\cite{bouchaud_universality_1997, majumdar_extreme_2020}. 
Hard constraints or boundaries, such as nonnegativity constraints in stochastic queues~\cite{lindley_theory_1952, boxma_multiplicative_2021} and price limits or circuit breakers in financial systems~\cite{abad_holding_2013}, can suppress extreme excursions.
When an observable is clipped at such a
constraint, the part of the fluctuation that would have carried it
farther is concealed from observation. We ask whether this hidden component merely disappears from observation
or can influence the subsequent evolution of the system.

To place this question in context, it is useful to recall how boundaries are conventionally represented in stochastic physics.
They delimit confined domains~\cite{chandrasekhar_stochastic_1943, guerin_mean_2016,  giuggioli_exact_2020}, define targets in diffusion and search processes~\cite{condamin_first_2005,benichou_first-passage_2014, chevalier_first-passage_2011, meyer_universality_2011, godec_proximity_2016, grebenkov_multi-target_2024}, represent reactive surfaces in diffusion-mediated phenomena~\cite{grebenkov_paradigm_2020, grebenkov_surface-hopping_2020}, and impose state-space constraints in reflected diffusion and stochastic networks~\cite{harrison_reflected_1981}. 
The resulting dynamics depend crucially on the rule applied at the boundary. An absorbing boundary terminates the trajectory and thereby defines relevant first-passage and persistence observables~\cite{redner_guide_2001,bray_persistence_2013,majumdar_universal_2010,majumdar_survival_2017,mattos_first_2012,pal_first_2019, das_discrete_2022, sarmiento_first-passage-time_2025, das_tethering_2026}.
Related applications include the study of a freezing transition in the barrier crossing rate~\cite{sabhapandit_freezing_2020}, extremal statistics of run-and-tumble particles~\cite{singh_extremal_2022}, and work extraction in quantum measurement engines~\cite{bresque_run-and-tumble_2025}.
A reflecting boundary keeps the process within the admissible region~\cite{angelani_run-and-tumble_2015, grebenkov_local-time_2019, metzler_boundary_2000, das_dynamics_2023}, whereas a partially reactive boundary permits repeated encounters before reaction, with finite surface reactivity controlling absorption~\cite{chaigneau_partially_2022,bressloff_switching_2022, guerin_imperfect_2023}. For nonlocal Lévy flights and finite-velocity Lévy walks, the choice
and implementation of boundary rules can lead to inequivalent dynamics
and to different survival, first-passage, and stationary statistics~\cite{dybiec_levy_2017,garbaczewski_fractional_2019}.

A different mechanism arises when the boundary acts through clipping. 
Although conventional boundary rules can themselves generate nontrivial 
temporal correlations, clipping has a distinct 
feature: different underlying excursions beyond the boundary are mapped 
onto the same observed boundary value. The clipped observable therefore 
loses information about the overshoot, while the underlying dynamics need 
not erase that information. If the concealed overshoot survives and is 
partially carried forward, information removed from the observable becomes 
a hidden dynamical state that influences subsequent evolution. Memory is 
then neither imposed through temporally correlated forcing nor attributed 
to a pre-existing slow variable; rather, clipping creates the hidden excess 
and retention converts it into a memory carrier. 
We ask whether this 
mechanism can generate memory even when the external forcing itself has 
no temporal dependence.

This question connects the problem of constrained fluctuations to the
broader study of memory in driven matter. Physical memory commonly
arises when threshold crossings, switching events, or irreversible
rearrangements modify an internal or configurational state
~\cite{keim_memory_2019}. In hysteretic systems, transitions between
metastable states preserve information about earlier extrema of the
applied drive~\cite{sethna_hysteresis_1993}. Cyclically driven
suspensions and amorphous materials can similarly encode previous
driving amplitudes through particle rearrangements and persistent
changes in microstructure
~\cite{keim_transient_2011,paulsen_multiple_2014,
galloway_structure_2022}, while interacting mechanical instabilities
can store information about past loading~\cite{shohat_memory_2022}. 
These examples illustrate a common principle: information absent from 
the instantaneous observable may remain stored in an internal state 
and affect later behavior. Our setting realizes this principle in a 
particularly minimal form, in which the internal state is not introduced 
independently but is precisely the excess concealed by clipping. Its 
partial retention couples successive time steps, while the hidden state 
vanishes whenever clipping does not occur.

Daily fixed price bands in financial markets provide a particularly
transparent setting for this retained-excess mechanism. 
In return units, a price band clips the observed daily return at an
exchange-imposed limit, while the concealed
component admits a natural interpretation as unresolved demand or
supply.
As the regulatory constraint acts during each trading
day, whereas the proposed retention acts from one market close to the
next, we distinguish the
discrete trading day index $t$ from the intraday time coordinate
$\tau\in[\tau_{\rm open},\tau_{\rm close}]$, where
$\tau_{\rm open}$ and $\tau_{\rm close}$ denote the opening and
closing times of the market, respectively.
Let $s_t(\tau)$ denote the price of a stock during
trading day $t$, with its closing price
$S_t=s_t(\tau_{\rm close})$. Relative to the previous-day closing 
price $S_{t-1}$, the intraday running return on day $t$ is
$r_t(\tau)=[s_t(\tau)-S_{t-1}]/S_{t-1}$, and the realized daily return
is its closing value 
$R_t=r_t(\tau_{\rm close})=(S_t-S_{t-1})/S_{t-1}$. Although $\tau$ is a
continuous intraday time coordinate, no continuity of the price path
$s_t(\tau)$ is assumed. The model developed in this work instead
advances in discrete time from one market close to the next.

Stock exchanges set daily price bands relative to the previous closing
price to restrain extreme price movements
~\cite{chan_price_2005,wong_magnet_2020}. 
During a trading day on which a fixed symmetric price band remains in
force, its representation in return units is the interval $[-C,C]$,
so that the running return satisfies
$-C\le r_t(\tau)\le C$. Here $C>0$ is the magnitude of the return 
limit, equivalently the half-width of the return band, while 
$+C$ and $-C$ are the upper and lower return limits.
Consequently, the realized closing return $R_t$ lies in $[-C,C]$. When 
the running return reaches either limit, the fixed band blocks further 
movement beyond it, although the return may subsequently move back into 
the interior of the band. If the running return is at either return limit 
when the market closes, then $R_t=C$ or $R_t=-C$. We refer to the events 
$R_t=C$ and $R_t=-C$ as upper- and lower-limit closes on day $t$, respectively, 
and collectively as limit closes; in market terminology, they are respectively 
known as upper and lower circuits. Thus, a price band moderates extreme 
observed movements but may also delay full price adjustment. This 
trade-off underlies the continuing debate over its effects on price discovery and 
market quality~\cite{chan_price_2005, kim_price_1997,berkman_effectiveness_2002}.

\begin{figure}[tbp]
    \centering
    \includegraphics[scale=0.8]{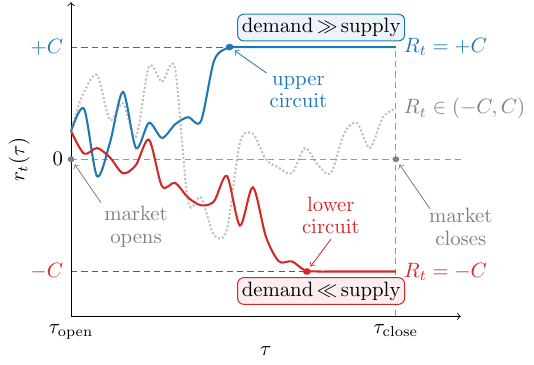}
    \caption{Possible intraday evolution of the running return
    $r_t(\tau)$ under a symmetric return band with limits at $\pm C$. The realized daily
    return is the closing value $R_t=r_t(\tau_{\rm close})$. A limit close records $R_t=\pm C$ but does not reveal the magnitude of the
    underlying excess demand or supply. Note that the discrete-time quantity $R_t$ does not resolve the intermediate intraday path; the model instead maps the state directly from one market close to the next.}
    \label{fig:schematic}
\end{figure}

A close at either return limit also entails a loss of information: the observed return $R_t$ no longer reveals how far the price would have moved in the absence of the limit.
Figure~\ref{fig:schematic} illustrates this loss. The curves show
possible intraday evolutions of the running return $r_t(\tau)$. When
the running return closes inside the return band, its closing value is
observed directly as $R_t$. 
In contrast, at a limit close, only the
value $R_t=\pm C$ is observed, while the strength of the market
pressure that would have moved the price further in the same direction
remains unknown. We refer to this unobserved part as the hidden excess. An
upper-limit close may therefore conceal unresolved excess demand,
whereas a lower-limit close may conceal unresolved excess supply.
Within this interpretation, a limit close indicates only the direction 
of the unresolved demand--supply imbalance, but not its magnitude.

We propose that part of the hidden excess is retained after the market closes and influences the return on the following trading day; clipping hides the excess, while retention converts that excess into memory. 
Thus, two limit closes with the same observed return can conceal different excesses and therefore carry different information about subsequent dynamics.
A same-sign next-day response can arise even when the fresh driving shocks are symmetric and independent across trading days.
This mechanism complements empirical studies documenting delayed price discovery, volatility spillovers, distortions in trading activity, order-book imbalances, and changes in liquidity or market quality around price-limit events~\cite{kim_price_1997,chan_price_2005, berkman_effectiveness_2002}. 
Here, we investigate how partial retention of the hidden excess can couple successive trading days. To describe this coupling quantitatively, we introduce a discrete-time latent variable and define the hidden excess as the part removed by clipping at each market close.

We formulate a minimal latent-state model and show that, for driving shocks with regularly varying tails, the stationary latent return preserves the shock tail index but has an enhanced tail amplitude. 
In the wide-band regime with large $C$, where limit closes become rare, such a close admits a single-dominant-shock description, consistent with the single-big-jump principle~\cite{vezzani_single-big-jump_2019,holl_big_2021}. 
The age and relative overshoot of this shock determine the leading-order next-day statistics. 
For nonzero retention, we derive finite nonzero limits for the normalized same-sign next-day mean return and the probability of a next-day close at the same return limit (\textit{persistence}), together with a power-law suppression of a next-day transition to the opposite return limit (\textit{reversal}). 
Numerical simulations support the asymptotic predictions, and data from the National Stock Exchange of India (NSE) show next-day same-sign responses across different price bands, consistent with theoretical predictions.

The paper is organized as follows.
Section~\ref{sec:latent-state-model} introduces the latent-state model
and the retention mechanism, while Sec.~\ref{sec:stationary-state}
establishes the stationary dynamics and its heavy-tailed behavior.
Section~\ref{sec:rare_circuit_histories} develops the single-dominant-shock description in the wide-band limit, and Sec.~\ref{sec:post_circuit_observables} derives the resulting next-day
mean response, the same-limit persistence probability, and the opposite-limit
reversal probability.
Section~\ref{sec:nse-comparison} compares the theoretical predictions
with the NSE data, and Sec.~\ref{sec:conclusion} concludes the article.
The appendices provide supporting derivations and additional details of
data construction and statistical estimation.

\section{Latent-state model}
\label{sec:latent-state-model}

We now formulate a discrete-time model for the retention of the hidden excess. 
The time is indexed by the trading day $t$, and all variables are defined at the close of the market. 
The model therefore describes the
evolution from one closing time to the next; it does not resolve the
intraday path $s_t(\tau)$ or the time at which either return limit is
reached.

To quantify the hidden component generated by clipping, let us first distinguish between the realized return observed in the market and an underlying latent return. The latent return, denoted by $X_t$, represents the unconstrained market response that would have been realized in the absence of the regulatory constraint. The realized return is obtained by clipping the latent return at the upper and lower return limits, i.e., 
\begin{align}
R_t=\operatorname{clip}(X_t,-C,C)=
\begin{cases}
-C, & X_t<-C,\\
X_t, & |X_t|\le C,\\
C, & X_t>C.
\end{cases}
\label{eq:clip_definition}
\end{align}
In the model, an upper-limit close is defined by $R_t=C$,
equivalently $X_t\ge C$, while a lower-limit close is defined by
$R_t=-C$, equivalently $X_t\le -C$. We refer to either event as a
limit close.

The clipping operation leaves part of the latent return unobserved. We therefore define the \emph{hidden excess}
\begin{align}
L_t=X_t-R_t,
\label{eq:hidden_excess}
\end{align}
which vanishes for $|X_t|\le C$ and is nonzero for $|X_t|>C$. The hidden excess thus represents the portion of the latent return removed from observation by clipping.

Our central hypothesis is that this hidden excess is not completely dissipated when trading ends. Instead, a fraction of it survives and contributes to the latent return on the following trading day according to
\begin{align}
X_{t+1}
=
\epsilon_{t+1}
+\lambda L_t,
\label{eq:latent_dynamics}
\end{align}
where $\{\epsilon_t\}$ is a sequence of i.i.d.~daily driving shocks, and the retention coefficient $\lambda$ (with $0 \le \lambda < 1 $) controls how much of the
hidden excess contributes to the next latent return.

Note that the hidden excess is the \emph{only} source of temporal coupling in the model. 
When no excess is removed by clipping, so that $|X_t|\le C$ and
$L_t=0$, the next latent return reduces to the independent driving
shock $\epsilon_{t+1}$. Increasing the retention coefficient $\lambda$ increases the contribution of the previous hidden
excess to the next latent return. 
The retention coefficient therefore controls the memory generated by the retained hidden excess.

Empirical financial returns exhibit strongly non-Gaussian,
heavy-tailed distributions over broad ranges of return magnitudes
~\cite{mantegna_scaling_1995,gopikrishnan_scaling_1999,
plerou_scaling_1999,cont_empirical_2001}. 
Here, we assume that driving shocks have a symmetric density,
$f(\epsilon)=f(-\epsilon)$, regularly varying tails, and a finite first
moment, without imposing a specific functional form.
A particular form of the distribution $f(\epsilon)$ is used later solely for numerical analysis.

Taking the conditional expectation of Eq.~\eqref{eq:latent_dynamics} yields
$\mathbb E[X_{t+1}\mid X_t] = \lambda L_t$,
since symmetry and the finite-first-moment assumption imply $\mathbb E[\epsilon_{t+1}]=0$. 
Thus, whenever clipping removes a nonzero excess
($|X_t|>C$, so that $L_t\neq0$), the latent return at time $t+1$ depends explicitly on the state at time $t$, even though the external driving shocks are independent. The hidden excess is therefore the unique source of temporal memory in the model.

We note that the model is formulated in terms of simple returns to match exchange-mandated price limits, which are defined as linear percentage deviations from the previous close. While the use of symmetric, unbounded driving shocks technically permits unphysical returns $X_t \le -1$, this boundary violation is negligible for realistic volatility scales. In this framework, the latent state $X_t$ serves as a continuous proxy for the unobserved market order imbalance rather than a direct price coordinate \cite{bouchaud_theory_2003}.

\section{Stationary state}
\label{sec:stationary-state}

Writing
$g(x)
\equiv
\lambda\!\left[
x-\operatorname{clip}(x,-C,C)
\right]$,
one has
$|g(x)-g(y)|\le\lambda|x-y|$ for $0\le \lambda<1$.
The latent dynamics~\eqref{eq:latent_dynamics} is therefore a contractive stochastic recursion. Under
the standard logarithmic moment condition
$\mathbb E[\log(1+|\epsilon_t|)]<\infty$, it admits a unique stationary
distribution~\cite{diaconis_iterated_1999}. 

Let $p(x)$ denote the stationary probability density of the latent return $X_t$.  
As a future latent return is obtained by adding a fresh shock to the retained hidden excess, Eq.~\eqref{eq:latent_dynamics} implies that conditioned on the present latent state $X_t=x$, the probability density of $X_{t+1}=x'$ is $\mathcal{K}(x'|x) = f\!\left(x'-g(x)\right)$, which defines the Markov transition kernel of the process.
Consequently, one has the self-consistent equation 
\begin{align}
p(x) = \int_{-\infty}^{\infty} p(y)\, f\!\left(x-g(y)\right) \,\dd y.
\label{eq:stationary_equation}
\end{align}
The stationary density inherits the symmetry of the driving shocks.
Indeed, since $f(\epsilon)=f(-\epsilon)$ and $g(-y)=-g(y)$, the
transformation $x\to-x$ and $y\to-y$ shows that $p(-x)$ also satisfies
Eq.~\eqref{eq:stationary_equation}. The uniqueness of the stationary
distribution therefore implies $p(x)=p(-x)$.

The stationary distribution of the realized return $R_t$ follows
immediately. 
Inside the return band $[-C,C]$, clipping is inactive, and the realized and latent returns coincide. The probability carried by latent states beyond the return limits is instead concentrated into point masses at the two values $\pm C$.

\begin{figure}[tbp]
    \centering
    \includegraphics[scale=1]{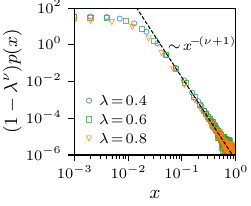}
    \caption{Positive-tail collapse of the stationary latent density $p(x)$ confirming the amplitude renormalization~\eqref{eq:stationary_tail} with $C=0.1$ and $\nu=3.5$. Symbols are obtained from stationary trajectories of length $10^{10}$ generated by numerical simulations using Student's $t$ driving shocks with degrees of freedom $\nu$ and scale parameter $s_{\epsilon} = 0.01$ [see Appendix~\ref{sec:student-t}].
    }
    \label{fig:stationary-tail}
\end{figure}

We assume that the symmetric driving shock density is regularly varying,
\begin{align}
f(\epsilon) \sim a|\epsilon|^{-(\nu+1)}, \qquad |\epsilon|\to\infty,
\label{eq:regular_variation}
\end{align}
where we refer to $\nu>1$ as the tail index; equivalently, the density has power-law exponent $\nu+1$. 
Assuming in addition that the stationary density $p(x)$ is ultimately monotone in each tail, in the far-tail limit at fixed return band~\cite{bingham_regular_1987}, it satisfies [see Appendix~\ref{sec:px-tail}],
\begin{align}
p(x)
\sim
\frac{a}{1-\lambda^\nu}
|x|^{-(\nu+1)},
\qquad
\frac{|x|}{C}\to\infty.
\label{eq:stationary_tail}
\end{align}
Retention of the hidden excess therefore preserves the shock tail
index while enhancing the stationary tail amplitude by the factor
$(1-\lambda^\nu)^{-1}$.
The predicted amplitude renormalization is verified numerically in
Fig.~\ref{fig:stationary-tail}. Multiplying the stationary density by
$(1-\lambda^\nu)$ collapses the positive tails obtained for different
values of $\lambda$ onto a common curve, confirming that retention
changes the asymptotic amplitude while leaving the power-law exponent unchanged.

\section{Dominant-shock histories in the wide-band limit}
\label{sec:rare_circuit_histories}

The stationary analysis above characterizes the far-tail regime
$|x|/C\to\infty$ at fixed $C$.
We now turn to the wide-band limit $C/s_\epsilon\to\infty$ at fixed
$\lambda$ and $\nu$, where $s_\epsilon$ denotes the characteristic scale
of the driving shocks.
Physically, this corresponds to $C\gg s_\epsilon$, so both the
upper and lower return limits lie far outside the typical shock scale,
and limit closes are rare.
Note that conditioning on an upper-limit close, i.e., $X_t\ge C$, selects latent states with
$X_t/C$ of order unity, rather than states asymptotically far beyond
the upper return limit.
The fixed-$C$ stationary-tail result~\eqref{eq:stationary_tail},
which applies when $|x|/C\to\infty$, therefore does not directly
describe this conditional regime and cannot be used directly to
calculate post-limit-close observables.
Instead, the single-big-jump principle for regularly varying shocks
motivates a single-dominant-shock description~\cite{vezzani_single-big-jump_2019,holl_big_2021}: an upper-limit close is
asymptotically generated by one dominant positive shock, which may
occur on day $t$ or be inherited from an earlier day through the
retained hidden excess.
By symmetry, it is sufficient to formulate the analysis for
upper-limit closes; the corresponding results for lower-limit closes
follow immediately with the sign reversal.

\subsection{Single-dominant-shock histories}

\begin{figure}[tbp]
    \centering
    \includegraphics[scale=1]{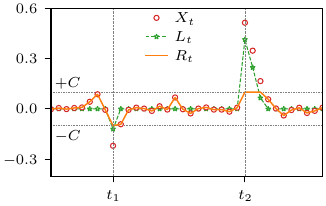}
    \caption{Representative discrete-time trajectory of the latent
    return $X_t$, realized return $R_t$, and hidden excess
    $L_t=X_t-R_t$. The large negative shock at $t_1$ produces a
    short-lived lower-limit close, whereas the large positive shock at
    $t_2$ generates a persistent hidden excess and consecutive
    upper-limit closes. Horizontal dotted lines mark the return limits
    $\pm C$, vertical dashed lines mark the shock times, and connecting
    lines are guides to the eye.}
    \label{fig:big_jump}
\end{figure}

Suppose that an upper-limit close is observed on trading day $t$, i.e.,
$X_t\ge C$. In the wide-band limit, typical driving shocks are negligible on
the scale $C$, and the observed close must be traced to an
exceptionally large positive shock. 
Figure~\ref{fig:big_jump} illustrates how large shocks can generate
different limit-close histories. On day $t_1$, a large negative shock
drives the latent return below the lower return limit, so that the
realized return is clipped at $R_{t_1}=-C$ and a negative hidden excess
is generated. This excess vanishes the following day, and the
process returns to the interior of the return band. On day $t_2$, by
contrast, a larger positive shock produces a substantial hidden
excess. Its retained part keeps the latent return above the upper
return limit for several subsequent days, resulting in consecutive
upper-limit closes. Thus, a limit close may be caused either by
a large shock on the same day or by the surviving excess generated by
an earlier shock. In particular, a sequence of limit closes need
not be produced by a sequence of large shocks. Histories containing two
or more shocks of order $C$ have a smaller asymptotic probability. In
the leading-order single-dominant-shock description, the dominant shock
is therefore explicitly retained, while the remaining shocks are
neglected on the scale $C$.

If $\lambda=0$, no hidden excess is carried forward, and an
upper-limit close on day $t$ can only be generated by the fresh shock
$\epsilon_t$. 
In the shock-age representation discussed below, this corresponds to
the $j=0$ history. We therefore develop the nontrivial retained-history
construction for $0<\lambda<1$.

Let the dominant shock responsible for the upper-limit close observed
on day $t$ have occurred $j$ trading days earlier, i.e., on day $t-j$,
where $j=0,1,2,\ldots$. 
In other words, the age of the dominant shock is $j$.
The case $j=0$ corresponds to a fresh shock that 
occurs on the day of the observed close, whereas $j\ge1$
corresponds to an earlier shock whose hidden excess survives until day
$t$. In leading order on the scale $C$, the latent state on the day of
the dominant shock is $X_{t-j}\simeq\epsilon_{t-j}$.

Consider a history in which the latent state remains above the upper
return limit from day $t-j$ to day $t$. On each such day,
$R_n=C$ and $L_n=X_n-C$. Since the nondominant shocks are neglected on
the scale $C$ in this leading-order description, the latent state
propagates according to
\begin{align}
X_{n+1}
\simeq
\lambda(X_n-C),
\qquad
n=t-j,\ldots,t-1.
\label{eq:largeC_deterministic_path}
\end{align}
Thus, clipping leaves the hidden excess $X_n-C$, of which only the
fraction $\lambda$ is transferred to the following day. Starting from
day $t-j$ and iterating
Eq.~\eqref{eq:largeC_deterministic_path} for $j$ steps gives
\begin{align}
X_t
\simeq
\lambda^j\epsilon_{t-j}
-
C\sum_{m=1}^{j}\lambda^m.
\label{eq:largeC_historical_path}
\end{align}
Imposing the observed upper-limit-close condition $X_t\ge C$ then
yields
\begin{align}
\epsilon_{t-j} \ge CB_j, \qquad B_j \equiv \sum_{m=0}^{j}\lambda^{-m}.
\label{eq:largeC_Bj}
\end{align}
The factor $B_j$ is therefore the minimum shock amplitude, measured in
units of $C$, required for a shock of age $j$ to survive repeated
clipping and retention and still produce the upper-limit close
observed on day $t$.

\subsection{Age distribution of the dominant shock}

Having determined the threshold associated with each shock age, we now calculate the relative statistical weights of the possible histories. Let
\begin{align}
\overline F_\epsilon(x) \equiv \Pr(\epsilon\ge x) \sim kx^{-\nu}, \qquad x\to\infty,
\label{eq:shock_survival_largeC}
\end{align}
denote the positive survival tail of the driving shocks, where $k=a/\nu$ follows from Eq.~\eqref{eq:regular_variation}.

For each $j\ge0$, define the event
\begin{align}
\mathcal A_j(C) \equiv \{\epsilon_{t-j}\ge CB_j\},
\end{align}
which states that the shock on day $t-j$ is large enough, by itself, to survive the $j$ clipping steps and generate the upper-limit close on day $t$. Its probability satisfies
\begin{align}
\Pr[\mathcal A_j(C)] = \overline F_\epsilon(CB_j) \sim kC^{-\nu}B_j^{-\nu}.
\label{eq:prob_A_j}
\end{align}
Although $j$ extends over the entire past, sufficiently old histories are strongly suppressed because $B_j^{-\nu}$ decreases geometrically with $j$ as $B_j^{-\nu}
\sim (1-\lambda)^\nu\lambda^{j\nu}$.

The family of events $\{\mathcal A_j(C)\}_{j\ge0}$ is not pairwise disjoint, since two historical shocks may exceed their respective thresholds. Such overlaps are asymptotically negligible. For $i\neq j$, the independence of shocks gives
\begin{align}
\Pr[ \mathcal A_i(C)\cap\mathcal A_j(C) ] \sim k^2C^{-2\nu}B_i^{-\nu}B_j^{-\nu}.
\label{eq:pair_overlap}
\end{align}
Since $\sum_{j\ge0}B_j^{-\nu}$ converges,  Eq.~\eqref{eq:pair_overlap} implies that the total probability of histories containing two or more dominant shocks is $O(C^{-2\nu})$, compared with $O(C^{-\nu})$ for histories containing a single dominant shock. Thus, the conditional probability of more than one order-$C$ shock vanishes in the wide-band limit.

Within this single-dominant-shock description, the upper-limit close on day $t$ is represented, to leading order, by the union of possible age-dependent histories, i.e., $\Pr(X_t\ge C) \sim \Pr\left( \cup_{j=0}^{\infty}\mathcal A_j(C) \right)$.
Because overlaps between these events are asymptotically negligible, we obtain
\begin{align}
\Pr(X_t\ge C) &\sim \sum_{j=0}^{\infty}\overline F_\epsilon(CB_j) \sim kC^{-\nu}\mathcal Z(\lambda,\nu),
\label{eq:upper_circuit_probability_largeC}
\end{align}
where the quantity 
\begin{align}
\mathcal Z(\lambda,\nu) \equiv \sum_{j=0}^{\infty}B_j^{-\nu}
\end{align}
is convergent.

Let $J$ denote the age variable of the dominant shock.
Equation~\eqref{eq:prob_A_j} implies that the age-$j$ contribution to
the upper-limit-close probability is asymptotically proportional to
$B_j^{-\nu}$. Conditioning on $X_t\ge C$ therefore gives
\begin{align}
\Pr(J=j\mid X_t\ge C)
\to
\pi_j(\lambda,\nu)
\equiv
\frac{B_j^{-\nu}}
{\mathcal Z(\lambda,\nu)}.
\label{eq:dominant_shock_age_distribution}
\end{align}
Thus, $\pi_j$ is the asymptotic probability that the upper-limit close
observed on day $t$ originated from a dominant shock that occurred $j$
trading days earlier. Older shocks carry smaller weights because they
must exceed progressively larger thresholds.

\subsection{Hidden excess conditional on shock age}

If a dominant shock of age $j$ produces the upper-limit close, its magnitude exceeds the threshold in Eq.~\eqref{eq:largeC_Bj}. 
Let us define its threshold-exceedance ratio by
\begin{align}
V \equiv \frac{\epsilon_{t-j}}{CB_j}, \quad V\ge1,
\label{eq:V-def}
\end{align}
for which the conditional survival probability in the wide-band limit is given by
$ \Pr(V>v\mid\mathcal A_j(C)) =
{\Pr(\epsilon_{t-j}>vCB_j) }/{ \Pr(\epsilon_{t-j}>CB_j) } 
\to v^{-\nu}$.
Thus, conditional on exceeding the threshold $CB_j$, the
threshold-exceedance ratio converges to a Pareto random variable $V$
with survival probability
\begin{align}
\Pr(V>v)=v^{-\nu}, \quad v\ge1,
\label{eq:pareto-survival-v}
\end{align}
and probability density
\begin{align}
\rho(v)=\nu v^{-(\nu+1)}, \quad v\ge1.
\end{align}
Importantly, this limiting distribution is independent of the shock
age $j$ and of the characteristic scale of the driving shocks.

From Eqs.~\eqref{eq:largeC_historical_path},~\eqref{eq:largeC_Bj}, and~\eqref{eq:V-def},  one may write the latent state on day $t$ as
\begin{align}
X_t \simeq C[1+A_j(V-1)], \quad A_j \equiv \sum_{m=0}^{j}\lambda^m = \lambda^jB_j.
\label{eq:Aj-def}
\end{align}
The corresponding hidden excess is
\begin{align}
L_t \simeq CA_j(V-1).
\label{eq:he-up-lim}
\end{align}
Here, the quantity $(V-1)$ is the relative overshoot above the minimum shock amplitude (see Eq.~\eqref{eq:V-def}) required to generate the present upper-limit close, while $A_j$
records how this overshoot is transformed along a history of $j$
retained clipping steps. 
Since $A_j$ increases with $j$, an older dominant shock with the same
relative overshoot $V-1$ leaves a larger hidden excess on day $t$.
This occurs because the threshold $CB_j$ required for an older shock
is itself larger.

\section{Post-limit-close observables}
\label{sec:post_circuit_observables}

The wide-band dominant-shock description developed above determines
the next-day mean response, the same-limit persistence probability,
and the opposite-limit reversal probability.

\subsection{Conditional mean response}

We define the conditional mean responses following upper- and
lower-limit closes as
\begin{align}
m_+(C)&\equiv\mathbb E[R_{t+1}\mid X_t\ge C],  \label{eq:mp-def} \\
m_-(C)&\equiv\mathbb E[R_{t+1}\mid X_t\le-C].
\end{align}
Conditional on an upper-limit close ($X_t \ge C$) generated by a dominant shock of
age $j$ and threshold-exceedance ratio $V$, the next-day latent return
satisfies
$
{X_{t+1}}/{C}
\simeq
{\epsilon_{t+1}} / {C}
+
\lambda A_j(V-1)
\simeq
\lambda A_j(V-1),
$
where we have used
$\epsilon_{t+1}/C\to0$ in probability because the fresh shock is
typically of order $s_\epsilon\ll C$ in the wide-band limit. Consequently, the next-day normalized
realized return, conditional on the shock age $j$, converges to
\begin{align}
\frac{R_{t+1}}{C} \to \min\left\{ \lambda A_j(V-1),1 \right\},
\label{eq:app_scaled_response_age_j}
\end{align}
where the minimum appears because of the upper clipping at
$R_{t+1}=C$. The lower return limit $-C$ does not contribute in leading order
because the retained shift is nonnegative and a fresh negative shock
of order $C$ has a probability that vanishes in the wide-band
limit.

Since the normalized realized return is bounded, the corresponding conditional 
limiting mean generated by
a shock of age $j$ is
\begin{align}
\mathbb E\left[ \frac{R_{t+1}}{C} \,\middle| X_t\ge C,\ J=j \right] \to H_j,
\label{eq:mean_conditional_j}
\end{align}
where
\begin{align}
H_j \equiv \mathbb E\left[ \min\{\lambda A_j(V-1),1\} \right].
\label{eq:app_H_definition}
\end{align}
For a random variable $Y$ that satisfies $0\le Y\le1$, one may use the
identity
$\mathbb E[Y] = \int_0^1\Pr(Y>y)\,\mathrm{d}y$.
Applying this identity to
$Y=\min\{\lambda A_j(V-1),1\}$ gives
\begin{align}
H_j =
\frac{\lambda A_j}{\nu-1} \left[ 1- \left( 1+\frac{1}{\lambda A_j} \right)^{1-\nu} \right],
\label{eq:app_H_closed}
\end{align}
where we have used the Pareto survival law for $V$ given in Eq.~\eqref{eq:pareto-survival-v}.

Within the wide-band single-dominant-shock description, an upper-limit close is an asymptotic mixture of the possible dominant-shock ages.
Averaging the age-dependent response $H_j$ with the conditional weights
$\pi_j(\lambda, \nu)$ and using Eqs.~\eqref{eq:mp-def} and~\eqref{eq:mean_conditional_j} gives the wide-band post-upper-limit-close response as ${m_+(C)} / {C} \to \sum_{j=0}^{\infty}\pi_j(\lambda,\nu)H_j$, which can be recast using Eq.~\eqref{eq:dominant_shock_age_distribution} as
\begin{align}
\frac{m_+(C)}{C}
\to \frac{1}{\mathcal Z(\lambda,\nu)}
\sum_{j=0}^{\infty}B_j^{-\nu}H_j \equiv
\mathcal M(\lambda,\nu).
\label{eq:full_largeC_response}
\end{align}
Within the class of regularly varying symmetric driving
distributions, the limiting normalized response~\eqref{eq:full_largeC_response} depends on the driving-shock distribution only through its tail index $\nu$ and on the dynamics only
through the retention coefficient $\lambda$. By symmetry, the corresponding
lower-limit-close response satisfies
${m_-(C)} /{C} \to -\mathcal M(\lambda,\nu)$.
For $\lambda=0$, no hidden excess is retained and
$\mathcal M(0,\nu)=0$, which can be interpreted as the continuous extension of
Eq.~\eqref{eq:full_largeC_response}.

\begin{figure}[tbp]
    \centering
    \includegraphics[scale=1.]{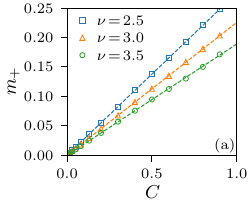}
    \includegraphics[scale=1.]{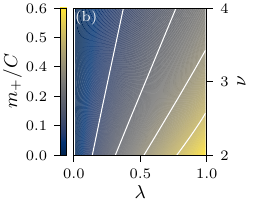}
    \caption{(a) Post-upper-limit-close response $m_+$ as a function of the return-limit magnitude $C$ for $\lambda=0.5$; dashed lines denote the analytical result~\eqref{eq:full_largeC_response}. Symbols are obtained from stationary trajectories of length $10^{10}$ generated by numerical simulations using Student's $t$ driving shocks with degrees of freedom $\nu$ and scale parameter $s_{\epsilon} = 0.01$ [see Appendix~\ref{sec:student-t}].  (b) Limiting normalized response $m_+/C$ in the $(\lambda,\nu)$ plane obtained from Eq.~\eqref{eq:full_largeC_response}.}
    \label{fig:mean-res}
\end{figure}

Figure~\ref{fig:mean-res}(a) supports the asymptotic scaling
$m_+(C)\sim C\mathcal M(\lambda,\nu)$. The response grows linearly with the return-limit magnitude $C$, with a
larger slope for heavier-tailed shocks.
Figure~\ref{fig:mean-res}(b) shows that the normalized response
increases with retention and is stronger for heavier-tailed shocks.
Stronger retention carries more hidden excess forward, while heavier
tails produce larger conditional overshoots above the age-dependent
threshold $CB_j$. The response is therefore the largest for strong
retention and heavy tails and vanishes only at $\lambda=0$.

Indeed, in the small-retention limit $\lambda\to0$, the $j=0$ contribution dominates because
$B_j^{-\nu}=O(\lambda^{j\nu})$ for $j\ge 1$.
Consequently,
$\mathcal M(\lambda,\nu)\sim{\lambda} /({\nu-1}),$
implying that there is no finite retention threshold: 
any nonzero
retained fraction produces a positive conditional mean return after an
upper-limit close and, by symmetry, a negative conditional mean return
after a lower-limit close.

\subsection{Persistence and reversal}

We next determine how an upper-limit close on day $t$ affects the
probability of a same-limit close or an opposite-limit close on day
$t+1$. We define
\begin{align}
P_{++}(C)
&\equiv
\Pr(X_{t+1}\ge C\mid X_t\ge C),
\\
P_{+-}(C)
&\equiv
\Pr(X_{t+1}\le-C\mid X_t\ge C).
\end{align}
The first quantity is the same-limit persistence probability, whereas
the second is the opposite-limit reversal probability.

Conditioned on a dominant shock of age $j$, the scaled hidden excess
satisfies $ {L_t}/{C} \simeq A_j(V-1)$; see Eq.~\eqref{eq:he-up-lim}.
Since a typical fresh next-day shock is negligible on the scale $C$,
the following day is an upper-limit close at leading order when
$
\lambda A_j(V-1)\ge1.
$
Using the Pareto survival law of $V$ therefore gives
\begin{align}
\Pr(X_{t+1} \!\ge\! C \!\mid\! J \!=\!j,X_t\ge C)
&\to \Big( 1+\frac{1}{\lambda A_j} \Big)^{-\nu} \!= \frac{\pi_{j+1}}{\pi_j},
\label{eq:per1}
\end{align}
where we have used $A_{j+1}=1+\lambda A_j$ and
$A_j=\lambda^jB_j$.

Equation~\eqref{eq:per1} has a simple interpretation in terms of the
shock-age distribution $\pi_j$. Conditioned on an upper-limit close on day $t$, the joint probability that the dominant shock has age $j$
and that the following day is also an upper-limit close satisfies
$\Pr(J=j,X_{t+1}\ge C\mid X_t\ge C)
=
\Pr(J\!=\!j \! \mid \! X_t\ge C)
\Pr(X_{t+1}\ge C \! \mid \! J \!=\! j,X_t\ge C)
\to 
\pi_{j+1}$.
Summing over all possible ages therefore gives
\begin{align}
P_{++}(C) \to
\sum_{j=0}^{\infty}\pi_{j+1}
=
1-\pi_0.
\label{eq:largeC_Ppp}
\end{align}
The limiting persistence probability is consequently equal to the
total statistical weight of histories with nonzero shock age.

\begin{figure}[tbp]
    \centering
    \includegraphics[scale=1.]{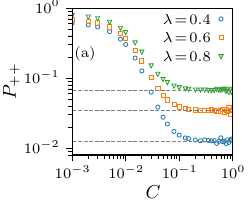}
    \includegraphics[scale=1]{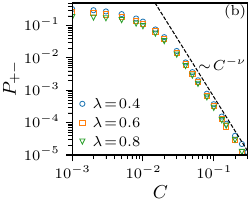}
    \caption{(a) Same-limit persistence probability $P_{++}(C)$ and
    (b) opposite-limit reversal probability $P_{+-}(C)$ as functions
    of the return-limit magnitude $C$ for $\nu=3.5$. Symbols are
    obtained from stationary trajectories of length $10^{10}$
    generated by numerical simulations using Student's $t$ driving
    shocks with $\nu$ degrees of freedom and scale parameter
    $s_{\epsilon}=0.01$ [see Appendix~\ref{sec:student-t}]. The
    horizontal dashed lines in (a) show the wide-band limit in
    Eq.~\eqref{eq:largeC_Ppp}, while the dashed line in (b) shows the
    $C^{-\nu}$ decay predicted by Eq.~\eqref{eq:largeC_Ppm}.}
    \label{fig:pers}
\end{figure}

For an opposite-limit reversal on day $t+1$, the fresh shock must
overcome both the lower return limit and the nonnegative retained
excess. Conditioned on shock age 
$J=j$ and $V=v$, this requires
$\epsilon_{t+1}
\le
-C\left[
1+\lambda A_j(v-1)
\right]$.
Using the symmetry relation $\Pr(\epsilon\leq-x)
=
\overline F_\epsilon(x)
\sim
kx^{-\nu}$, and the regular variation of the shock tails, one obtains
$
\Pr(
X_{t+1}\leq-C
\mid J=j,V=v,X_t\geq C)
\sim
kC^{-\nu}
\left[
1+\lambda A_j(v-1)
\right]^{-\nu}.
$
Averaging over $V$ and the shock-age distribution gives
\begin{align}
P_{+-}(C)
\sim
kC^{-\nu}\Psi(\lambda,\nu),
\label{eq:largeC_Ppm}
\end{align}
where
\begin{align}
\Psi(\lambda,\nu)
&\equiv 
\frac{1}{\mathcal Z(\lambda, \nu)} \sum_{j=0}^{\infty} B_j^{-\nu} \int_1^\infty  \frac{ \nu v^{-(\nu+1)} }{ \left[ 1+\lambda A_j(v-1) \right]^\nu } \,\dd v \nonumber \\
&=\frac{1}{2\mathcal Z(\lambda,\nu)}\sum_{j=0}^{\infty}B_j^{-\nu}{}_2F_1\left(\nu,1;2\nu+1;1-\lambda A_j\right),
\label{eq:Psi_definition}
\end{align}
with ${}_2F_1$ being the hypergeometric function.

Such a reversal requires the dominant positive shock responsible
for the initial upper-limit close and an additional independent
negative shock of order $C$ on the following day. The joint occurrence
of these two extreme shocks has probability of order $C^{-2\nu}$.
Since the probability of the initial upper-limit close is of order
$C^{-\nu}$, conditioning on that close leaves
$P_{+-}(C)=O(C^{-\nu})$. For fixed nonzero retention, the retained
hidden excess therefore produces an $O(1)$ same-limit persistence
probability, whereas an opposite-limit reversal requires a second
extreme fluctuation and is asymptotically suppressed.
This asymptotic separation is verified numerically in
Fig.~\ref{fig:pers}, which shows that $P_{++}(C)$ approaches the finite limit
$1-\pi_0$, while $P_{+-}(C)$ decays as $C^{-\nu}$.

\section{Comparison with stock-market data}
\label{sec:nse-comparison}

We now compare theoretical predictions with real market data by analyzing upper- and lower-limit closes in NSE stocks between November 2007 and June 2026. Return-limit magnitudes are fixed by NSE at $C=2\%,\,5\%,\,10\%$ and $20\%$. 
Appendix~\ref{sec:data} describes the details of data processing. 
Both free parameters entering the theoretical comparison, the tail index
$\nu$ and the retention coefficient $\lambda$, are fixed without
fitting the empirical post-limit-close mean responses.

We fix the tail index at $\nu=3$, the representative heavy-tail
benchmark selected in Appendix~\ref{sec:nu-estimation} from the
standardized daily returns of 13 major global equity indices. This
value is obtained entirely independently of the NSE limit-close data
analyzed in the following.

The retention coefficient is estimated from the same-limit persistence probability at the widest return band $C=20\%$; the narrower bands exceed the maximum persistence allowed by the wide-band asymptote and therefore cannot be used for this inversion. 
At $C=20\%$, 217 of the 1494 upper- and lower-limit closes are followed
by another close at the same return limit. 
Using Eq.~\eqref{eq:largeC_Ppp}, we solve
$1 - \pi_0(\lambda_{\rm eff},3) = 1-1/\mathcal Z(\lambda_{\rm eff},3)=217/1494$ and obtain
$\lambda_{\rm eff}\approx 0.942$.
The complete persistence counts and the admissibility test are given
in Appendix~\ref{sec:lambda-estimation}. 
Since the hidden excess is not
directly observable and can vary across stocks and events,
$\lambda_{\rm eff}$ should be interpreted as an effective aggregate
parameter rather than as a microscopic retention fraction.

\begin{figure}[tbp]
    \centering
    \includegraphics[scale=0.95]{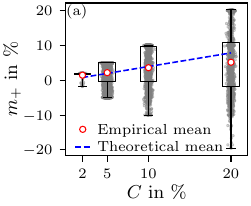}
    \includegraphics[scale=0.95]{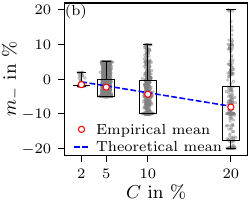}
\caption{Next-day returns following NSE limit closes from
November 2007 to June 2026, for return-limit magnitudes
$C=2\%,\,5\%,\,10\%$, and $20\%$. Box plots summarize the empirical
distributions (box: 25th--75th percentiles; whiskers: minimum to
maximum), jittered points show individual observations, and red
circles mark the empirical means. (a) Next-day return following an
upper-limit close. (b) Next-day return following a lower-limit close.
The blue dashed line in (a) shows the wide-band asymptotic prediction
$m_+(C)\sim C\mathcal M(\lambda_{\rm eff},\nu)$, with $\nu=3$ and
$\lambda_{\rm eff}=0.942$; the line in (b) shows
$m_-(C)\sim-C\mathcal M(\lambda_{\rm eff},\nu)$ using the same
parameters.}
    \label{fig:nse}
\end{figure}

With $\nu=3$ and $\lambda_{\rm eff}=0.942$ fixed as described above,
Fig.~\ref{fig:nse} compares the theoretical responses $m_+(C)$ and
$m_-(C)$ from Eq.~\eqref{eq:full_largeC_response} with the measured
post-limit-close returns. 
Let $\widehat m_+(C)$ and $\widehat m_-(C)$ denote the empirical
conditional mean next-day returns following upper- and lower-limit
closes, respectively.
At every return-limit magnitude $C$, the empirical conditional means have
the signs predicted by the model: upper-limit closes are followed, on
average, by a positive next-day return, whereas lower-limit closes are
followed by a negative one. 
However, the magnitudes of the empirical mean responses are not exactly symmetric, so $\widehat m_+(C) \neq - \widehat m_-(C)$.
Within the present symmetric formulation, the upper- and lower-limit responses are mirror images, so the modest directional asymmetry observed empirically provides a natural motivation for extensions with direction-dependent retention or asymmetric driving-shock statistics.

The theory captures the overall scale and the increase in the mean response
magnitude with the return-limit magnitude $C$. 
The agreement is closest near $C=10\%$.
The largest relative discrepancy occurs in the narrowest band,
$C=2\%$, consistent with the failure of the wide-band persistence
criterion in this band. At $C=20\%$, the empirical upper-limit-close
response lies substantially below the theoretical prediction,
producing the largest absolute discrepancy in that direction.

Taken together, the data are consistent with the qualitative
predictions of a same-sign post-limit-close response and an increasing
response magnitude with the return-limit magnitude $C$.
However, quantitatively, a single effective
retention coefficient does not reproduce the complete band and
the directional dependence of the empirical means. 
This directional asymmetry reflects a well-known feature of financial markets: extreme negative shocks are often driven by systemic panic, margin calls, and forced liquidations, which breed further mechanical selling~\cite{bouchaud_leverage_2001}. In contrast, extreme positive shocks are typically driven by idiosyncratic news that is rapidly priced in. Consequently, downward limits generate a stronger structural memory than upward limits, which explains why the empirical lower-limit persistence at the 20\% band (see Appendix~\ref{sec:lambda-estimation}) exceeds the theoretical upper bound of the symmetric i.i.d. model.

\section{Conclusion}
\label{sec:conclusion}

We have introduced a minimal discrete-time model in which temporal memory is generated by partial retention of the hidden excess produced by clipping. Clipping alone limits the realized return and conceals the excess beyond the return limits, but it does not create temporal dependence. Memory arises only when a fraction of this excess is carried into the latent return on the following trading day. In this sense, the memory is carried by information that clipping removes from the instantaneous observable but retention preserves in the dynamics. Accordingly, in the absence of retention, the latent dynamics reduces to the independent driving shocks.

For regularly varying driving shocks, the model admits a unique stationary
state whose latent-return distribution preserves the shock tail index
while exhibiting a retention-dependent enhancement of the far-tail
amplitude. In the wide-band limit, a limit-close may be generated either by
a large shock on the same day or by the surviving hidden excess of a
shock from an earlier day. The age and relative overshoot of this
dominant shock determine the leading next-day statistics.

This mechanism produces a marked asymmetry between persistence and
reversal. In the wide-band limit, we obtain a nonzero normalized
same-sign mean response for fixed nonzero retention and a same-limit
persistence probability $P_{++}(C)\to1-\pi_0$. In contrast, an opposite-limit reversal requires an additional extreme shock and is suppressed algebraically,
$P_{+-}(C)\sim kC^{-\nu}\Psi(\lambda,\nu)$. Numerical simulations
support these asymptotic predictions. The NSE data are consistent with
the predicted same-sign response and its increase with the return-limit
magnitude, although a single effective retention coefficient does not
reproduce the full dependence on the return band and on the direction
of the initial limit close.

The empirical directional asymmetry suggests several extensions.
Different retention coefficients for upper- and lower-limit closes
could describe different relaxation rates of unresolved excess demand
and excess supply. Other natural directions include asymmetric driving
shocks, heterogeneity across stocks and events, and finite-band
corrections beyond the wide-band limit. More generally, the
persistence--reversal asymmetry provides a testable signature of
retained hidden excess in systems where clipping or saturation conceals
part of a fluctuation while allowing that concealed component to remain
dynamically active.

\section*{Acknowledgments}
The author acknowledges support from the MUR PRIN2022 project ``Breakdown of ergodicity in classical and quantum many-body systems'' (BECQuMB) Grant No. 20222BHC9Z.

\section*{Data availability}
The data supporting the plots and results within this paper are available from the author upon reasonable request.

\appendix
\section{Asymptotic tail of the latent process}
\label{sec:px-tail}

To determine the stationary tail of the latent process $X_t$, it is
convenient to consider its survival probability,
$\overline P(x)=\Pr(X_t>x)$. When $\lambda=0$, the dynamics reduces to
$X_t=\epsilon_t$, so the stationary tail coincides directly with the
shock tail. We therefore focus below on the nontrivial retention regime
$0<\lambda<1$.

We first verify that the latent dynamics belongs to the class of
contractive stochastic recursions with regularly varying inputs
considered in Ref.~\cite{buraczewski_asymptotics_2012}. One may recast 
Eq.~\eqref{eq:latent_dynamics} as
\begin{align}
X_{t+1}
=
\Phi_{\epsilon_{t+1}}(X_t),
\end{align}
where
\begin{align}
\Phi_{\epsilon}(x)
&\equiv
\epsilon+\lambda h(x),
&
h(x)
&\equiv
x-\operatorname{clip}(x,-C,C).
\end{align}
The hidden-excess map $h$ has the piecewise-linear form
\begin{align}
h(x)
=
\begin{cases}
x+C, & x<-C,\\
0,   & -C\le x\le C,\\
x-C, & x>C.
\end{cases}
\end{align}
It is therefore continuous and nondecreasing, with slope either zero
or one. More explicitly, for $x<y$, one has 
$ 0 \le h(y)-h(x) \le y-x$,
and hence, for arbitrary $x,y\in\mathbb R$,
\begin{align}
|h(x)-h(y)|
\le
|x-y|.
\label{eq:h_lipschitz}
\end{align}
Accordingly, for every fixed $\epsilon$,
\begin{align}
\left|
\Phi_{\epsilon}(x)-\Phi_{\epsilon}(y)
\right|
&=
\lambda|h(x)-h(y)|
\nonumber\\
&\le
\lambda|x-y|.
\label{eq:update_contraction}
\end{align}
Thus, the recursion is globally contractive with contraction
coefficient $\lambda<1$.
To make this explicit, let
$X_{t+n}^{(x)}$ and $X_{t+n}^{(y)}$ denote two trajectories initialized
at $X_t^{(x)}=x$ and $X_t^{(y)}=y$, respectively, and driven by the
same subsequent shock sequence. Iterating
Eq.~\eqref{eq:update_contraction} over $n$ time steps gives
\begin{align}
\left|
X_{t+n}^{(x)}-X_{t+n}^{(y)}
\right|
\le
\lambda^n|x-y|,
\end{align}
so the recursion is geometrically contractive in every positive moment order.

The update map may also be written as
$ \Phi_{\epsilon}(x) = \epsilon+\lambda x+r(x)$, where  $r(x) \equiv -\lambda\operatorname{clip}(x,-C,C)$ denotes a bounded state-dependent correction to the linear
map $\epsilon+\lambda x$, satisfying
$\sup_{x\in\mathbb R}|r(x)| \le \lambda C$.
In the notation of Ref.~\cite{buraczewski_asymptotics_2012}, the
regularly varying component is therefore the shock $\epsilon$, the
linear coefficient is the deterministic constant $\lambda$, and the
remaining state-dependent component is uniformly bounded. The
required contractivity and moment conditions are consequently
satisfied. It follows that the stationary distribution is regularly
varying with the same tail index $\nu$ as the driving shocks. We may
therefore write
\begin{align}
\overline P(x)
\sim
K_+x^{-\nu},
\qquad
x\to\infty,
\label{eq:tail_ansatz}
\end{align}
where $K_+$ is an unknown amplitude.

At stationarity, Eq.~\eqref{eq:latent_dynamics} gives
\begin{align}
\overline P(x)
=
\Pr\!\left(
\epsilon_{t+1}+\lambda L_t>x
\right).
\end{align}
Since $L_t$ is determined by the shocks up to day $t$, it is
independent of the fresh shock $\epsilon_{t+1}$. 
The shock assumption gives regularly varying positive and negative
tails for $\epsilon_{t+1}$ with index $\nu$. The tail
ansatz~\eqref{eq:tail_ansatz}, together with symmetry and the relation
between $L_t$ and $X_t$, gives the same tail index for
$\lambda L_t$. Since the two variables are independent, the
convolution-tail property of regularly varying random variables~\cite{embrechts_subexponentiality_1979, embrechts_convolution_1982} gives
\begin{align}
\Pr\!\left(
\epsilon_{t+1}+\lambda L_t>x
\right)
\sim
\Pr(\epsilon_{t+1}>x)
+
\Pr(\lambda L_t>x).
\label{eq:stationary_tail_convolution}
\end{align}
The fresh-shock contribution is
$\Pr(\epsilon_{t+1}>x) \sim kx^{-\nu}$, where 
$k\equiv {a} / {\nu}$; see
Eq.~\eqref{eq:regular_variation}.
On the other hand, for $x>0$, the event $\{\lambda L_t>x\}$ necessarily lies in the region $X_t>C$, where $L_t=X_t-C$. Therefore, we have 
\begin{align}
\Pr(\lambda L_t>x) &= \Pr\left( X_t>\frac{x}{\lambda}+C \right) = \overline P \left(\frac{x}{\lambda}+C \right) \nonumber\\
&\sim K_+ \left( \frac{x}{\lambda}+C \right)^{-\nu} \nonumber\\
&\sim K_+\lambda^\nu x^{-\nu},
\label{eq:retained_excess_tail}
\end{align}
where the fixed shift $C$ is asymptotically negligible as
$x\to\infty$.

Combining Eqs.~\eqref{eq:tail_ansatz},
\eqref{eq:stationary_tail_convolution}, and
\eqref{eq:retained_excess_tail}, and matching the stationary tail
amplitudes, gives
$ K_+ = k+\lambda^\nu K_+$, and hence
\begin{align}
K_+
=
\frac{k}{1-\lambda^\nu},
\end{align}
and therefore
\begin{align}
\overline P(x)
\sim
\frac{k}{1-\lambda^\nu}
x^{-\nu},
\qquad
\frac{x}{C}\to\infty
\quad\text{at fixed }C.
\label{eq:latent_survival_tail}
\end{align}
This expression also recovers the no-retention case $\lambda=0$.

If, in addition, the stationary density $p(x)$ is ultimately monotone
in each tail, the monotone-density theorem
~\cite{bingham_regular_1987}, together with Eq.~\eqref{eq:latent_survival_tail}, yields
\begin{align}
p(x)
\sim
\nu K_+x^{-(\nu+1)}
=
\frac{a}{1-\lambda^\nu}x^{-(\nu+1)}.
\end{align}
By symmetry, the same asymptotic holds in the negative tail, giving
Eq.~\eqref{eq:stationary_tail} of the main text.

\section{The shock distribution}
\label{sec:student-t}

For the numerical simulations, we sample the driving shock $\epsilon$ from a symmetric Student's $t$ distribution with degrees of freedom $\nu$ and scale parameter $s_{\epsilon}$ and density function given by
\begin{align}
\label{eq:student-t-distr}
f(\epsilon; \nu, s_{\epsilon}) = \frac{K_\nu}{s_{\epsilon}} \Big(1 + \frac{\epsilon^2}{\nu s_{\epsilon}^2}\Big)^{\!\!-\frac{\nu+1}{2}}, 
\end{align}
where 
\begin{align}
K_\nu = \frac{\Gamma\left(\frac{\nu+1}{2}\right)}{\sqrt{\nu\pi} \, \Gamma\left(\frac{\nu}{2}\right)}.
\end{align} 
The large-$|\epsilon|$ asymptotic form is
\begin{align}
f(\epsilon; \nu, s_{\epsilon})
\sim a |\epsilon|^{-(\nu+1)},
\qquad |\epsilon|\to\infty,
\label{eq:student_t_density_tail}
\end{align}
where $a\equiv K_\nu \nu^{(\nu+1)/2}s_\epsilon^\nu$.
The corresponding survival tail in the limit $x\to\infty$ reads
$\Pr(\epsilon>x) =
\int_x^\infty f(u)\,\mathrm{d}u
\sim a \nu^{-1} x^{-\nu}$, which yields symmetric tail probabilities, namely, 
$\Pr(\epsilon>x)\sim k\,x^{-\nu}$ and 
$\Pr(\epsilon<-x)\sim k\,x^{-\nu}$,
where $k\equiv a / \nu $.
Substituting these amplitudes into
Eq.~\eqref{eq:latent_survival_tail} gives the stationary latent-tail
amplitude used in the numerical comparison.

\section{Data processing and sample construction}
\label{sec:data}

In this appendix, we describe how the raw exchange data were turned into the empirical sample analyzed in Sec.~\ref{sec:nse-comparison}.
Here, we use the market terms \emph{upper circuit} and \emph{lower circuit} for the empirical counterparts of the upper- and lower-limit closes defined in the model. 
We refer to either case as a \emph{circuit event}.
We first describe the data sources and the rule used to identify a circuit event, then the exclusion criteria applied before computing the response and persistence statistics, and the resulting sample sizes.

\subsection{Data sources and circuit identification}
\label{sec:data:circuit}

Daily open, high, low, close, volume, and previous-close values for
equities listed on the National Stock Exchange (NSE) of India were
obtained from NSE's public data archives using the \texttt{nselib} Python package~\cite{nselib_python}.
The raw data reported by the exchange are not adjusted for corporate actions such as stock splits and bonus issues. The closing price on
trading day $t$ is denoted $S_t$, as in the Introduction. NSE
periodically revises the percentage price band (``circuit limit'')
applicable to each equity and discloses each such revision as a
discrete event, recording the band value in effect immediately before
and immediately after the change (or an explicit \emph{no-band} state,
in which no daily price limit applies). These disclosures were
consolidated into a single per-symbol chronological record that covered 
their full available history, beginning 31 October 2007.
Corporate-action disclosures -- restricted to bonus issues and stock
splits, the two action types with a determinate, exchange-disclosed
price ratio -- were likewise obtained for every symbol in the analysis
universe, recording the ex-date (the first trading session from which
the security trades without entitlement to the action) and the terms
of the action.

A symbol's day-by-day band history can be reconstructed unambiguously
only if its full recorded sequence of band-change events is internally
consistent: the state following one disclosed event must equal the
state preceding the symbol's next disclosed event. The history of every symbol was checked against this criterion, and a symbol failing at
any point was excluded from the analysis universe in its entirety,
leaving a base universe of $N_{\mathrm{sym}} = 757$ symbols.

For a trading day on which a numeric price band $C$ (in \%) is
known to be in force, the applicable upper and lower price limits are
\begin{equation}
  U_t = S_{t-1}\left(1 + \frac{C}{100}\right), \quad
  D_t = S_{t-1}\left(1 - \frac{C}{100}\right),
  \label{eq:limits}
\end{equation}
where $S_{t-1}$ is the closing price of the previous session, taken directly
from the exchange record without recomputation. The trading day $t$ is
classified as an \emph{upper-circuit} day if
$S_t \geq U_t(1-\tau)$, and as a \emph{lower-circuit} day if
$S_t \leq D_t(1+\tau)$, with tolerance $\tau = 0.25\%$. This criterion
is one-sided: no upper bound is placed on how far $S_t$ may exceed
$U_t$ (respectively, fall below $D_t$). This choice reflects a
documented mechanism in the data--the exchange occasionally revises
a stock's price band intraday after the nominal limit is first reached,
permitting continued trading beyond it--and a two-sided or
exact-match criterion would misclassify such exchange-sanctioned
trading as a data anomaly.

\subsection{Exclusion criteria and derived statistics}
\label{sec:data:filters}

Starting from the base universe, daily observations were excluded
according to the following criteria, applied in sequence:
(1) \textit{Low-price exclusion.} Observations with $S_t$ or $S_{t-1}$
    below INR~10 were excluded, removing a deep-penny-stock regime in
    which several of the largest apparent band overshoots were traced
    to stale, long-unrevised band records rather than genuine circuit
    events.
(2) \textit{Trading continuity.} Observations for which the next
    available trading record for that symbol occurred more than four
    calendar days later, or for which no subsequent record existed,
    were excluded, removing cases spanning a trading suspension or
    relisting rather than a consecutive session.
(3) \textit{Fixed band.} Observations for which the price band
    recorded on the following trading day differed from that recorded
    on the day of the circuit event were excluded, consistent with the
    constant-band assumption, as required by the theory.
(4) \textit{Corporate actions.} NSE's previous-close field is not
    adjusted for corporate actions; this was verified directly against
    a disclosed bonus issue, for which the reported previous close on
    the ex-date equals the unadjusted prior close rather than a
    share-basis-adjusted value. Observations were therefore excluded
    whenever a disclosed bonus issue or stock split's ex-date -- mapped
    to the affected symbol's own next actual trading day, to account
    for concurrent suspensions -- coincided with either the
    observation day or the immediately following trading day.
(5) \textit{Residual band exceedance.} Observations for which the
    realized return on event day, or on the following day,
    exceeded the nominal band by more than an additional tolerance
    $0.01\,C$ were excluded. This criterion is independent of and
    additional to the tolerance $\tau$ of
    Sec.~\ref{sec:data:circuit}: $\tau$ sets the floor for classifying
    a day as a circuit event at all, while this final criterion sets
    the ceiling on how far a same- or next-day return may extend past
    $C$ before being attributed to an unresolved band revision or
    residual stale-record artifact rather than a clean single-band
    event.
Table~\ref{tab:funnel} summarizes the effect of each stage on the
pooled sample of daily observations.

\begin{table}[tbp]
  \centering
  \caption{Sample size after each successive exclusion criterion.}
  \label{tab:funnel}
  \begin{tabular}{lr}
    \hline\hline
    Stage & Observations \\
    \hline
    Initial (symbol/date-range selection)      & 1{,}163{,}004 \\
    After low-price exclusion                & 1{,}046{,}516 \\
    After trading-continuity exclusion         & 1{,}040{,}234 \\
    After fixed-band exclusion                 &   922{,}417 \\
    After corporate-action exclusion           &   922{,}219 \\
    After residual band-exceedance exclusion   & {898{,}689} \\
    \hline\hline
  \end{tabular}
\end{table}

For each nominal band $C \in \{2\%, 5\%, 10\%, 20\%\}$ and each circuit
direction, the following quantities were calculated from the final
analysis sample:
\begin{align}
  \widehat m_+(C) &= \left\langle R_{t+1} \,\middle|\, \text{upper circuit at } t \right\rangle, \\
  \widehat m_-(C) &= \left\langle R_{t+1} \,\middle|\, \text{lower circuit at } t \right\rangle,
\end{align}
where $R_{t+1}$ is the close-to-close return on the trading day
following the circuit event, together with the \emph{persistence
probability} (the empirical frequency with which the same circuit
direction recurs on the following day) and the \emph{reversal
probability} (the empirical frequency with which the opposite circuit
direction occurs on the following day). 
The empirical results are compared with the
prediction~\eqref{eq:full_largeC_response} of our theoretical model in Fig.~\ref{fig:nse}.

\section{Selection of a representative heavy-tail benchmark}
\label{sec:nu-estimation}

To select the heavy-tail index $\nu$ independently of stock-level circuit
events, we use Yahoo!~Finance~\cite{yahoofinance} to analyze the historical daily closing prices $S_{i,t}$ of 13
major stock market indices between January 1990 and June 2026. The start dates
differ across the indices according to data availability. For index $i$,
we define the relative daily return
$
R_{i,t}= {(S_{i,t}-S_{i,t-1})} / {S_{i,t-1}}
$
and the standardized return
\begin{align}
Z_{i,t}=\frac{R_{i,t}-\mu_{R,i}}{\sigma_{R,i}}.
\end{align}
Each standardized series therefore has zero sample mean and unit sample
variance.
For each index $i$, we model the empirical probability density function (PDF) of standardized returns using a centered Student's $t$ distribution $f(Z_i; \nu_i, s_{Z,i})$ given in Eq.~\eqref{eq:student-t-distr} and estimate the parameters $\nu_i$ and $s_{Z,i}$ by maximum likelihood.
For a Student's $t$-distributed variable $X$ with scale parameter $s_X$ and degrees of freedom 
$\nu>2$, the variance is
$\mathrm{Var}(X)=s_X^2\nu/(\nu-2)$. An exact unit-variance Student's $t$ distribution would therefore have $s_{Z,i}^{\mathrm{th}}=\sqrt{(\nu_i-2)/\nu_i}$.
As $Z_i$ is obtained from empirical data, the finite-sample
maximum-likelihood estimate $s_{Z,i}$ need not equal
$s_{Z,i}^{\mathrm{th}}$ exactly.

\begin{table*}[htbp]
\centering
\caption{Empirical return statistics, fitted Student's $t$ distribution parameters, and the derived unstandardized scale parameters across global stock market indices.}
\label{tab:student_t_fits}
\begin{tabular}{l @{\hspace{15pt}} c @{\hspace{15pt}} c@{\hspace{5pt}}c c @{\hspace{15pt}} ccc  c @{\hspace{15pt}} c}
\hline\hline
 & & \multicolumn{2}{c}{Empirical} & & \multicolumn{3}{c}{Student's $t$ Fit} & & Unstandardized Scale \\
\cline{3-4} \cline{6-8} 
Index & $N$ & $\mu_{R,i}$ (\%) & $\sigma_{R,i}$ (\%) & & ${\nu}_i$ & $s_{Z,i}$ & KS $D_i$ & & $s_{R,i}$ (\%) \\
\hline
  S\&P 500     & 9188 & 0.039 & 1.13 & & 2.83 & 0.600 & 0.026 & & 0.681 \\
  Nasdaq 100   & 9188 & 0.067 & 1.66 & & 2.97 & 0.630 & 0.029 & & 1.048 \\
  Dow Jones    & 8682 & 0.038 & 1.09 & & 2.92 & 0.604 & 0.023 & & 0.661 \\
  FTSE 100     & 9215 & 0.022 & 1.07 & & 3.38 & 0.663 & 0.016 & & 0.712 \\
  DAX          & 9230 & 0.038 & 1.37 & & 3.32 & 0.665 & 0.023 & & 0.911 \\
  CAC 40       & 9224 & 0.025 & 1.33 & & 3.62 & 0.688 & 0.015 & & 0.915 \\
  Nifty 50     & 4604 & 0.045 & 1.30 & & 2.86 & 0.593 & 0.015 & & 0.771 \\
  SENSEX       & 7139 & 0.050 & 1.41 & & 3.14 & 0.639 & 0.017 & & 0.904 \\
  Nikkei 225   & 8952 & 0.018 & 1.49 & & 3.99 & 0.715 & 0.014 & & 1.064 \\
  KOSPI        & 7271 & 0.048 & 1.68 & & 2.41 & 0.554 & 0.026 & & 0.932 \\
  Hang Seng    & 9002 & 0.035 & 1.56 & & 3.37 & 0.660 & 0.015 & & 1.027 \\
  IBOVESPA     & 8214 & 0.130 & 2.12 & & 3.16 & 0.626 & 0.011 & & 1.324 \\
  TAIEX        & 7102 & 0.032 & 1.36 & & 3.04 & 0.647 & 0.023 & & 0.880 \\
  \hline
  {Average} & --- & {0.045} & {1.43} & &   {3.15} & {0.637} & --- & &   {0.910} \\
\hline\hline
\end{tabular}
\end{table*}

Table~\ref{tab:student_t_fits} shows the empirical statistics and the estimated parameters for each stock market index.
To quantify the overall distributional agreement for each index $i$, we compute the corresponding 
Kolmogorov--Smirnov distance
$
D_i=\sup_z\left|F_N(Z)-F(Z)\right| ,
$
where $F_N(Z)$ denotes the empirical cumulative distribution function, while $F(Z)$ is the analytical Student's $t$ CDF.
Across the 13 indices, $D_i\le0.029$, so the maximum absolute discrepancy
between the empirical and fitted cumulative distributions is below
$2.9$ percentage points, indicating close
agreement at the level of the full CDF.
Once the parameter ${s}_{Z,i}$ is estimated for a given index $i$, the unstandardized scale parameter $s_{R,i}$ in the original return units is obtained by post-hoc rescaling $s_{R,i} = {s}_{Z,i}  \sigma_{R,i}$.

\begin{figure*}[htbp]
    \centering
    \includegraphics[scale=0.9]{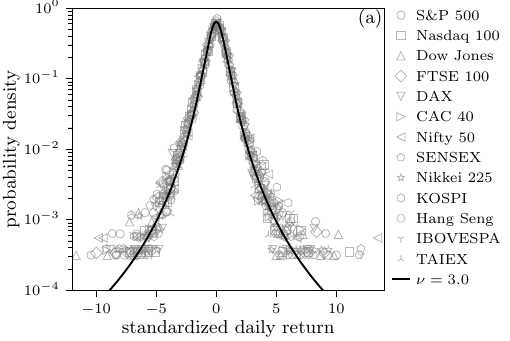}\hspace{10pt}
    \includegraphics[scale=0.9]{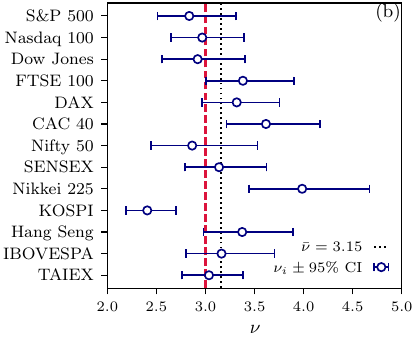}
\caption{Cross-market robustness of the heavy-tail benchmark for 13
global equity indices over their available histories between 1990 and
2026.
(a) Empirical probability densities of standardized daily returns
$Z=(R-\mu_R)/\sigma_R$. The black curve is the unit-variance Student's
$t$ reference distribution with $\nu=3$ and
$s_Z=1/\sqrt{3}$. Its survival tail obeys
$\Pr(|Z|>z) \propto z^{-3}$.
(b) Maximum-likelihood estimates $\nu_i$ with $95\%$
moving-block bootstrap confidence intervals obtained from
$B=2000$ resamples with block length $b=20$ trading days. The red
dashed line marks the inverse-cubic benchmark $\nu=3$, and the black
dotted line marks the unweighted cross-index mean
$\bar\nu=3.15$.}
    \label{fig:index_collapse}
\end{figure*}

Figure~\ref{fig:index_collapse} assesses the cross-market robustness of
the heavy-tail benchmark. Despite substantial differences in geography
and daily volatility (measured in terms of $\sigma_{R,i}$), the standardized return distributions show an
approximate collapse onto a common non-Gaussian form with pronounced
heavy tails; see Fig.~\ref{fig:index_collapse}(a). 
The fitted degrees-of-freedom estimates $\nu_i$ range from $2.41$ to $3.99$, with an unweighted
cross-index mean $\bar\nu=3.15$. The black curve in
Fig.~\ref{fig:index_collapse}(a) is the unit-variance Student's $t$
reference distribution with $\nu=3$, which we use as a representative
heavy-tail benchmark selected independently of the stock-level circuit
events.

To quantify the sampling uncertainty in the maximum-likelihood estimates
$\nu_i$, Fig.~\ref{fig:index_collapse}(b) reports $95\%$ moving-block-bootstrap
confidence intervals obtained from $B=2000$ resamples with block length
$b=20$ trading days. The moving-block bootstrap preserves temporal dependence over lags
contained within each 20-day block. The estimates exhibit measurable cross-market variation
but are collectively centered near the benchmark $\nu=3$.

\section{Estimation of the effective retention coefficient}
\label{sec:lambda-estimation}

We estimate the effective retention coefficient from the probability that a circuit event is followed by another circuit event at the same boundary. For each band $C$, let $N_+(C)$ and $N_-(C)$ denote the
numbers of upper- and lower-circuit events with a usable next-day
observation. The corresponding numbers followed by another circuit at
the same boundary are denoted by $N_{++}(C)$ and $N_{--}(C)$. The
directional persistence probabilities are given by
\begin{align}
P_{++}^{\rm emp}(C) &=N_{++}(C)/N_+(C) , \\
P_{--}^{\rm emp}(C) &=N_{--}(C)/N_-(C).
\end{align}
Since the asymptotic theory predicts the same large-$C$ persistence probability for
both upper and lower boundaries, we define a pooled estimator 
\begin{align}
P_{\rm pers}^{\rm emp}(C)
=
\frac{N_{++}(C)+N_{--}(C)}
     {N_+(C)+N_-(C)}.
\label{eq:app_pooled_persistence}
\end{align}
The measured counts and probabilities are reported in
Table~\ref{tab:persistence_counts}. The directional estimates are not
identical, particularly at the wider bands. The pooled value should
therefore be viewed as a model-based aggregate estimator, rather than
as empirical confirmation of exact upper--lower symmetry.

\begin{table}[tbp]
\centering
\caption{Empirical counts and probabilities of same-boundary
persistence.}
\label{tab:persistence_counts}
\begin{tabular}{c c c c c c c c}
\hline\hline
$C$ & $N_{+}$ & $N_{++}$ & $P_{++}^{\rm emp}$ & $N_{-}$ & $N_{--}$ & $P_{--}^{\rm emp}$ & $P_{\rm pers}^{\rm emp}$ \\
\hline
$2\%$  & 83    & 62   & 0.7470 & 144  & 117  & 0.8125 & 0.7885 \\
$5\%$  & 11129 & 5153 & 0.4630 & 8696 & 3724 & 0.4282 & 0.4478 \\
$10\%$ & 1542  & 387  & 0.2510 & 623  & 192  & 0.3082 & 0.2674 \\
$20\%$ & 1230  & 163  & 0.1325 & 264  & 54   & 0.2045 & 0.1452 \\
\hline\hline
\end{tabular}
\end{table}

For fixed $\nu$, the asymptotic persistence probability of Eq.~\eqref{eq:largeC_Ppp} equals 
$Q(\lambda,\nu) \equiv 1 - \pi_0 = 1-1/\mathcal Z(\lambda,\nu)$, which increases
monotonically with $\lambda$. As $\lambda\to1^-$,
$B_j\to j+1$ and $\mathcal Z(\lambda,\nu)\to\zeta(\nu)$, where $\zeta(\nu)$ denotes the Riemann zeta function, so that the function 
$Q$ is bounded above by
$Q_{\max}(\nu)=1-1/\zeta(\nu)$. For $\nu=3$, one thus has an upper bound $Q_{\max}(3) \approx 0.1681$.

As seen in Table~\ref{tab:persistence_counts}, the pooled persistence probabilities at $C=2\%,\,5\%$ and $10\%$
exceed this bound and therefore cannot be described by the large-$C$
asymptote for any admissible value of $\lambda$. 
The $20\%$ band is the only band for which the pooled persistence lies
within the range allowed by the large-$C$ asymptote. We therefore use
this pooled value to define an effective aggregate retention coefficient. 
Conditional on the choice of the benchmark $\nu=3$, the numerical solution of $Q(\lambda_{\rm eff},3)=P_{\rm pers}^{\rm emp}(20\%)$ gives
$\lambda_{\rm eff}\approx 0.942$, which is used in the main text.
We note that although the pooled persistence at $C=20\%$ is admissible
under the wide-band formula, the lower-limit estimate $P^{\mathrm{emp}}_{--}$ considered separately remains above the theoretical ceiling. The inferred
$\lambda_{\rm eff}$ is therefore an event-weighted pooled calibration,
not a direction-resolved fit.


\end{document}